\documentstyle[aps,epsf]{revtex}
\input{epsf}
\begin{document}

\draft

\title{Cooling the atomic motion with quantum interference}

\author{Giovanna Morigi} 

\address{Abteilung Quantenphysik, Universit\"at Ulm,
  Albert-Einstein-Allee 11, D-89081 Ulm, Germany
}
\date{\today}
\maketitle

\begin{abstract}
We theoretically investigate the quantum dynamics of the center of mass of
trapped 
atoms, whose internal degrees of freedom are driven in a $\Lambda$-shaped
configuration with the lasers tuned at two-photon resonance. In
the Lamb-Dicke regime, when the motional wave packet is well localized over
the laser wavelenght, transient coherent population 
trapping occurs, cancelling transitions at the laser frequency. In this limit
the motion can be efficiently cooled to the ground state of the trapping
potential. We derive an equation for the center-of-mass motion by
adiabatically eliminating the internal degrees of freedom. This treatment 
provides the theoretical background of the scheme presented in [G. Morigi
{\it et al}, Phys. Rev. Lett. {\bf 85}, 4458 (2000)] and implemented in
[C.F. Roos {\it et al}, Phys. Rev. Lett. {\bf 85}, 5547 (2000)]. We discuss
the physical mechanisms determining the dynamics and identify new
parameters regimes, where cooling is efficient. We discuss implementations of
the scheme to cases where the trapping potential is not harmonic. 
\end{abstract}

%\PACS{32.80.Pj,42.50.Gy,42.50.Vk}

\section{Introduction}

The progress in laser cooling of atoms and ions has set the stage
for coherent control of the dynamics of quantum mechanical
systems~\cite{Metcalf}. By means of laser cooling,
states of the center-of-mass motion of trapped atoms with high purity have
been prepared~\cite{Metcalf,SidebandIons,RamanIons,RamanAtoms}, 
allowing for instance for their coherent manipulation for quantum information
processing~\cite{QC}. Nevertheless, there is
a continuous interest for new and efficient cooling methods, which 
solve experimental difficulties and increase the efficiency of the process. 
In this context, a laser-cooling scheme for trapped atoms
has been recently proposed \cite{EITtheo2000}, that exploits the principles of
Coherent Population Trapping (CPT)~\cite{CPT} 
and allows to achieve almost unit probability of occupation of the
trapping-potential ground state \cite{EITtheo2000,EITexp2000}. 
This method has been
demonstrated to be an alternative to sideband~\cite{SidebandIons} and
Raman-sideband cooling~\cite{RamanIons,RamanAtoms}, 
routinely used for the preparation of very pure states of the center-of-mass
motion of trapped atoms and ions. 
Further applications of this cooling method
(now labeled as "EIT cooling") has been discussed in several
publications~\cite{ApplB2001,Hansch2001}.  

The focus of this work is to discuss theoretically the physical principles on
which this method is based, and particularly the role of quantum coherence 
between atomic states on the mechanical effects of light on trapped atoms. 
Thus, in the first section we introduce the electronic level scheme
composed by two stable or metastable states coupled by lasers to a common
excited state, the $\Lambda$ configuration, 
and discuss in general CPT when the
transitions are driven by counterpropagating laser beams (Doppler-sensitive
case). Here, we observe that in presence of an external potential confining
the center-of-mass motion, (transient) CPT is obtained when the lasers are set
at two-photon resonance and the wave packet is
well localized over the laser wavelength (Lamb-Dicke regime).
In the second section, starting from a general approach we develop the
theoretical model, assuming that the atomic center of mass is confined by an
external potential in the Lamb-Dicke regime: That allows to 
adiabatically eliminate the internal degrees of freedom and derive 
an equation for the external degrees of freedom only~\cite{Lindberg84}. 
We discuss this equation in detail
when the potential is harmonic, and derive a set of rate
equations for the occupation of the vibrational states. Thereby, we 
identify the parameters regime where cooling is effective. In some limits,
these equations reduce to the ones used
in~\cite{EITtheo2000,EITexp2000,ApplB2001}.
Nevertheless, a result of this paper is the identification of the basic
mechanism characterizing the dynamics, that allows us 
to determine new parameter 
regimes where cooling can be efficient. We discuss the
limit of validity of the equations derived, give alternative interpretations
of the dynamics, and consider possible extensions of the method
to cases, where the center of mass
is confined by a potential, that is not necessarily harmonic and whose
functional form may depend on the electronic state.

We remark that the laser-cooling dynamics of trapped atoms, whose internal
transitions are driven in a $\Lambda$ configuration, have been 
investigated in several works, as for instance
in~\cite{Lindberg86,Wineland92,Marzoli94,Dum94}. These, however, 
focused on different cooling mechanisms. This work, together
with~\cite{EITtheo2000}, extends these previous
analyses to other regimes, characterized by novel features of the 
center-of-mass dynamics, as we discuss below.

\section{The dark resonance and the motion}

In this section, we first discuss the internal dynamics and steady state
of an atom whose electronic bound states are driven by lasers in a resulting
$\Lambda$-configuration. We focus on the conditions for which CPT occurs. 
Then, we consider the 
center-of-mass degrees of freedom and discuss under which conditions
the features characterizing the bare internal dynamics are preserved, when the
motion is taken into account.
The discussion in this section and throughout the paper is
restricted to motion in one dimension, identified here with the
$\hat{x}$-axis. This allows a simpler exposition without loss of generality.
\begin{center}
\begin{figure}
\epsfxsize=0.4\textheight
\epsffile{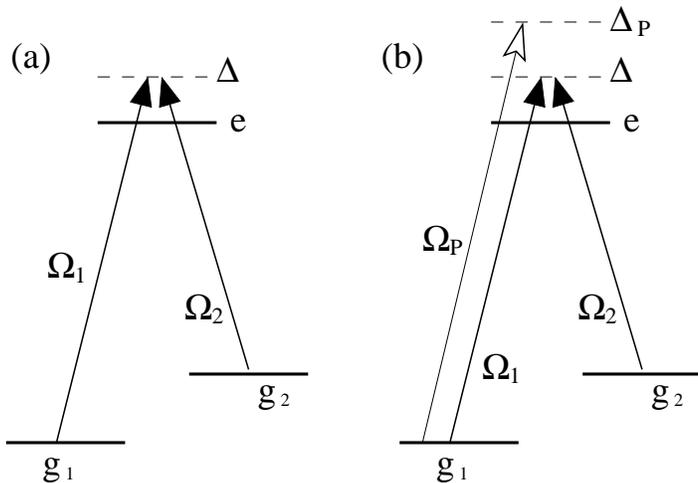}
\caption{(a) Level Scheme: The solid arrows represent the 
lasers at Rabi frequencies $\Omega_1$, $\Omega_2$, that couple to the
  transitions $|g_1\rangle\to|e\rangle$, $|g_2\rangle\to|e\rangle$,
  respectively, and are detuned of $\Delta$ from atomic resonance. 
(b) Addition of a probe at Rabi frequency $\Omega_{\rm P}$ and detuning 
$\Delta_{\rm P}$, coupling $|g_1\rangle\to|e\rangle$.}
\end{figure}
\end{center}

\subsection{The dark resonance}

An exemplary atomic level configuration where the effects of quantum
interference manifest is the $\Lambda$-transition. It consists
in two electronic transitions, formed by two stable or metastable states 
that we label $|g_1\rangle$, $|g_2\rangle$, which are coupled  by lasers
to the same excited state $|e\rangle$. For a closed transition, 
the atom stops to fluoresce when the states $|g_1\rangle$ and $|g_2\rangle$ are
resonantly coupled (two-photon resonance), as shown in Fig.~1(a): 
The system evolves into the dark state, a stable atomic-states superposition
which is decoupled from the excited state because of destructive
interference between the excitation amplitudes. This phenomenon is called
Coherent Population Trapping~\cite{CPT}, 
and the atoms are found in the coherence (dark state)  
\begin{equation}
|\Psi_{\rm D}\rangle = \frac{1}{\Omega}
 \left(\Omega_2|g_1\rangle-\Omega_1|g_2\rangle\right), \label{Dark}
\end{equation}
where $\Omega=\sqrt{\Omega_1^2+\Omega_2^2}$ and 
$\Omega_1$ ($\Omega_2$) is the Rabi frequency of the laser coupling to
the transition $|g_1\rangle\to|e\rangle$ ($|g_2\rangle\to|e\rangle$). 
Here, without loss of generality we have assumed $\Omega_1$, $\Omega_2$ real.
The dark state is accessed by spontaneous 
emission, unless the system has been initially prepared in it.
Thus, the density matrix $\rho_{\rm D}=|\Psi_{\rm D}\rangle 
\langle \Psi_{\rm D}|$ is the steady-state solution of the master equation for
the atomic density matrix $\rho$: $\partial\rho/\partial t={\cal
  L}_0\rho$, where ${\cal L}_0$ is the Liouvillian defined as  
\begin{equation}
\label{Master} {\cal L}_0\rho=\frac{1}{{\rm
    i}\hbar}[H,\rho]+{\cal K}\rho.
\end{equation}
Here, $H=H_0+V_0$ is the Hamilton operator, and its terms have the
form (in the rotating wave approximation and in the frame rotating at
the laser frequencies)
\begin{eqnarray}
\label{H:1} H_0&=&-\hbar\Delta \left(|g_1\rangle\langle g_1| 
+|g_2\rangle\langle g_2|\right),\\ 
\label{V:1} V_0&=&\frac{\hbar}{2} \left(
  \Omega_1 |e\rangle\langle g_1|+\Omega_2 |e\rangle\langle g_2|
+{\rm H.c.}\right),
\end{eqnarray}
where $\Delta=\omega_1-\omega_{\rm L,1}=\omega_{2}-\omega_{\rm
  L,2}$ are the laser detunings, with the atomic
  resonance frequencies  $\omega_{j}$ of the transition  
$|g_j\rangle\to|e\rangle$ and the
frequencies of the corresponding driving laser $\omega_{{\rm L},j}$
($j=1,2$). The operator ${\cal K}$ is the Liouvillian describing 
spontaneous emission, 
\begin{equation}
\label{K:1} {\cal K}\rho=-\frac{\gamma}{2} \left[|e\rangle\langle e|\rho+\rho
  |e\rangle\langle e|\right] +\sum_{j=1,2}\gamma_j |g_j\rangle\langle
e|\rho|e\rangle\langle g_j|, 
\end{equation}
where $\gamma_1$, $\gamma_2$ are the rate of decay into $|g_1\rangle$,
$|g_2\rangle$, respectively, and $\gamma_1+\gamma_2=\gamma$. It can be easily
verified that the dark state is a dressed state of the 
system, i.e. an eigenstate of $H$. The other two dressed states 
read~\cite{AtomPhoton} 
\begin{eqnarray}
\label{Dr:1}
&&|\psi_+\rangle=\cos\theta|e\rangle+\sin\theta|\psi_{\rm C}\rangle,\\
&&|\psi_-\rangle=\sin\theta|e\rangle-\cos\theta|\psi_{\rm C}\rangle,
\label{Dr:2}
\end{eqnarray}
where
\begin{eqnarray}
&&\tan\theta=\frac{\sqrt{\Delta^2+\Omega^2}-\Delta}{\Omega},\\
&&|\psi_{\rm C}\rangle=\frac{1}{\Omega}\left(\Omega_1|g_1\rangle
+\Omega_2|g_2\rangle\right),
\end{eqnarray}
and where we have introduced the state $|\psi_{\rm C}\rangle$, 
orthogonal to $|e\rangle$ and $|\psi_{\rm D}\rangle$. 
The states (\ref{Dr:1}), (\ref{Dr:2}) 
are at eigenfrequencies $\delta\omega_{\pm} = (\Delta\mp
\sqrt{\Delta^2+\Omega^2})/2$, and since they possess 
non-zero overlap with 
the excited state $|e\rangle$, they have a finite decay rate and 
are populated in the transient dynamics. We denote their linewidths 
with $\gamma_+$, $\gamma_-$.
The steady state is accessed at the slowest rate of decay
and, for later convenience, we introduce $T_0$, the time
scale corresponding to the inverse of this rate.\\
The dressed-state picture is a useful tool 
for interpreting the atomic spectra in a pump-probe experiment, where,
e.g., a weak probe at Rabi frequency $\Omega_{\rm P}$ ($\Omega_{\rm P}\ll
\Omega_1,\Omega_2$) couples to the transition $|g_1\rangle\to|e\rangle$ as
shown in Fig.~1(b), while
its frequency is let sweep across the atomic resonance. Figure~2 shows the
spectrum of excitation as a function of the detuning of the probe $\Delta_{\rm
  P}$, for a certain choice of the lasers parameters. Here, 
one can observe that
the component of the spectrum at $\Delta_{\rm P}=\Delta$ is zero, 
corresponding to the situation where the system is in the dark state
$|\Psi_{\rm D}\rangle$. Moreover, the spectrum 
exhibits two resonances centered at $\Delta_{\rm P}=\delta\omega_{\pm}$, 
whose widths correspond approximately (when
$|\Delta|,\Omega\gg\gamma$) to $\gamma_+$, $\gamma_-$, respectively, and can
be identified with the dressed states $|\psi_+\rangle$,
$|\psi_-\rangle$~\cite{Lounis92}. Note that these resonance have not a 
Lorentzian shape: The
spectrum shares in fact many similarities with a Fano profile~\cite{Lounis92}. 
Typical excitation spectra, measured with a
single ion in a trap, are reported in~\cite{Janik85,Stalgies98}.
\begin{center}
\begin{figure}
\epsfxsize=0.3\textheight
\epsffile{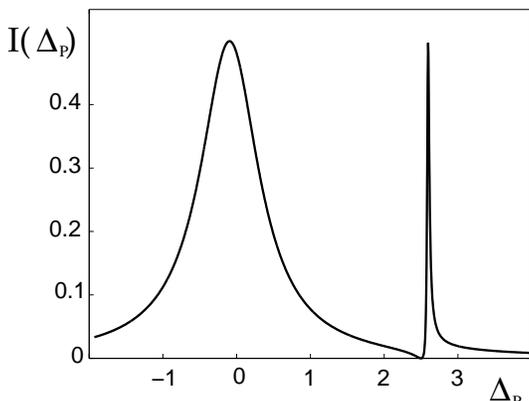}
\caption{Excitation spectrum $I(\Delta_{\rm P})$ in arbitrary units 
as a function of the probe detuning
$\Delta_{\rm P}$ in unit of $\gamma$. 
Here, $\Omega=\gamma$, $\Delta=2.5\gamma$
$\Omega_{\rm P}=0.05\gamma$.} 
\end{figure}
\end{center}
\subsection{The motion}

We consider now the center-of-mass motion in presence of a conservative
potential, of which for the moment the form is not specified. Given the
mass of the atom $m$, the momentum $p$, the position $x$ and the potential
$U(x)$, the mechanical Hamiltonian is 
\begin{equation}
H_{\rm mec}=\frac{p^2}{2m}+U(x).
\end{equation} 
We denote with $|\psi_{\epsilon}\rangle$ the eigenvectors of 
$H_{\rm mec}$ at the
eigenvalues $\epsilon$. The full dynamics are now described by the Master
equation
\begin{equation}
\label{Master:1} \frac{\partial}{\partial t}\tilde{\rho}=\frac{1}{{\rm
    i}\hbar}[\tilde{H},\tilde{\rho}]+\tilde{\cal K}\tilde{\rho} =\tilde{\cal
    L}\tilde{\rho},
\end{equation}
where $\tilde{\rho}$ is the density matrix for the internal and external
degrees of freedom and
\begin{equation}
\tilde{H}=\tilde{H}_0+H_{\rm mec}+\tilde{V}.
\end{equation}
Here, $\tilde{V}$ describes the coherent interaction of the
atomic dipole with the lasers, and has the form  
\begin{equation}
\label{Vmec} \tilde{V}=
\frac{\hbar}{2} \left( \Omega_1 {\rm e}^{{\rm i} k_1x\cos\phi_1}
  |e\rangle\langle g_1|+\Omega_2 {\rm e}^{{\rm i} k_2x\cos\phi_2}
   |e\rangle\langle g_2|+{\rm H.c.}\right),
\end{equation}
where the lasers are travelling waves at wave vectors $k_1$ and $k_2$, 
propagating along the directions forming the angles $\phi_1$, $\phi_2$,
respectively, with the $\hat{x}$-axis. In~(\ref{Vmec}) 
the spatial dependence is explicitly included, which couples to the external
degrees of  freedom of the ion, while the Rabi frequencies $\Omega_1$,
$\Omega_2$  are assumed to be constant over the spatial region where the ion
is localized.  
The liouvillian $\tilde{\cal K}$ describes the incoherent scattering
processes, whereby a photon is spontaneously emitted under an angle $\phi$
with the axis of the motion. It has the form:
\begin{eqnarray}
\label{Kmec} 
\tilde{\cal K}\tilde{\rho}=
&-&\frac{\gamma}{2} \left[|e\rangle\langle e|\tilde{\rho}+\tilde{\rho}
  |e\rangle\langle e|\right]\\
&+&\sum_{j=1,2}\gamma_j \int_{-1}^1 {\rm
  d}\cos\phi {\cal N}(\cos\phi) |g_j\rangle\langle e|\left[{\rm e}^{{\rm
      i}k_jx\cos\phi}\tilde{\rho}{\rm e}^{-{\rm i}k_jx\cos\phi}\right]
|e\rangle\langle g_j|, \nonumber
\end{eqnarray}
where ${\cal N}(\cos\phi)$ is the probability distribution for the angles of
photon emission respect to the motional axis.\\ In this system, at a given
instant of time perfect destructive interference between excitation amplitudes
occurs for the state 
\begin{equation}
|\tilde{\Psi}_{\rm
 D}\rangle=\frac{1}{\Omega}\left(\Omega_2|g_1,\Psi \rangle 
- \Omega_1{\rm e}^{{\rm i}(k_1\cos\phi_1-k_2\cos\phi_2)x} 
|g_2,\Psi \rangle\right),\label{Dark:1}
\end{equation}
where $\exp({\rm i}(k_1\cos\phi_1-k_2\cos\phi_2)x)$ 
is the displacement operator, acting on the
external degrees of freedom, and $\Psi$ is a
state of the center-of-mass motion. The state~(\ref{Dark:1}) is stable -and
thus a dark state- if it is eigenstate of $H_0+H_{\rm mec}$. 
This is always true when the lasers are copropagating and
$k_1\cos\phi_1=k_2\cos\phi_2$  (or, for one-dimensional motion as in this
case, when the direction of propagation of the lasers is orthogonal to the axis
of the motion, $\cos\phi_1=\cos\phi_2=0$): Then,
the motional state factorizes out in (\ref{Dark:1}). 
For $k_1\cos\phi_1\neq k_2\cos\phi_2$, on the  
contrary, one must consider the particular form of the confining
potential. For instance, for free atoms ($U(x)=$ const.) a perfect dark state
exists for $k_1\cos\phi_1=-k_2\cos\phi_2=k$ and reads $|\tilde{\Psi}_{\rm
  D}\rangle =\left(\Omega_2|g_1,-\hbar k\rangle -\Omega_1|g_2,\hbar
  k\rangle\right)/\Omega$. This property has been used to prepare very cold
atomic samples~\cite{VSCPT}. In presence of a confining
potential, on the other hand, there exists in general no state
$|\tilde{\Psi}_{\rm D}\rangle$ that is perfectly dark. Approximate dark
states have been discussed in~\cite{Dum94} for a 1D flat-bottom and for a 2D
harmonic trap. 

Nevertheless, transient
CPT can be observed in trapping potentials
and in Doppler-sensitive configurations when the atoms are in the 
Lamb-Dicke regime (LDR), i.e. 
when the size of their motional wave packet $\sqrt{\langle \Delta x^2\rangle}$
is much smaller than the wavelength of the
light, $k_{1,2}\sqrt{\langle \Delta x^2\rangle}\ll 1$.
In this limit, a hierarchy of processes in the excitation of the
center-of-mass wave packet is estabilished. At
zero order in $\zeta=k_{1,2}\sqrt{\langle \Delta x^2\rangle}$,
the effects due 
to the spatial gradient of the light-atom potential are neglected: 
the atoms behave like they were point-like, and the coherent
transitions take place at the laser frequency (carrier). Then, after the
transient time $T_0$ the atoms have
accessed the internal dark state $|\Psi_{\rm D}\rangle$. At first order in
$\zeta$, effects due to the finite size of the motional wave-packet become
manifest, and transitions between different motional states (sidebands
transitions) occur. On this longer time scale, that we denote with 
$T_{\zeta}$, the atom is optically pumped out of the dark state into another
state of the motion. In the Lamb-Dicke regime, the relation $T_{\zeta}\gg T_0$
allows for 
a coarse-grained description of the dynamics, where the internal state
of the atom is assumed to be always the dark state $|\Psi_{\rm
  D}\rangle$. \\
\indent 
These arguments suggest that for a  
trapped atom in the LDR some of the properties of the excitation spectrum 
discussed for the Doppler-free
case may also be applicable to the Doppler-sensitive one. 
Here, the carrier transition is predominant, whereas transitions
which change the state of the motion (sidebands transitions) are of higher
order in the Lamb-Dicke parameter, and can be interpreted as transitions due
to a probe ($\Omega_{\rm P}$)
set at the corresponding frequency in the bare atom, as illustrated for
instance in Fig.~1(b). In the next 
section, we show that this interpretation is theoretically justified.

\section{Theory}

Here, we derive the equations for the center-of-mass motion in the limit 
where the LDR applies and when the center-of-mass motion is confined by the
same potential at all three electronic levels. The procedure
consists in  adiabatically eliminating the internal degrees of freedom
from the dynamical equation at second order in the parameter $\zeta$,
and it corresponds to analysing the coarse-grained
evolution on the time
interval $\Delta t$ such that $T_{\zeta}\gg \Delta t\gg
T_0$. The formalism we use 
has been first developed in~\cite{Lindberg84} for a two-level transition
driven by a running wave, and later applied to standing-wave drives and
multilevel transitions in~\cite{Cirac92}. In the following, we 
outline the fundamental steps that are most general to all treatments, 
and refer the reader to~\cite{Lindberg84,Cirac92} for details 
(we have used the same notation as in~\cite{Cirac92} when possible).

\subsubsection{Lamb-Dicke limit}

In the Lamb-Dicke limit $\zeta\ll 1$, the operators $\exp({\rm i}k_jx)$
appearing in (\ref{Vmec}),(\ref{Kmec}) can be expanded in powers of
$\zeta$. At second order in this
expansion Eq.~(\ref{Master:1}) can be rewritten as
\begin{equation}
\frac{\partial}{\partial t}\tilde{\rho}
=\left[\tilde{\cal L}_0+\tilde{\cal L}_1
+\tilde{\cal L}_2\right]\tilde{\rho},
\end{equation}
where the Liouvillians $\tilde{\cal L}_j$ describe processes 
at the $j$th order in the Lamb-Dicke parameter, and are defined as:
\begin{eqnarray}
\tilde{\cal L}_0\tilde{\rho}
&=&{\cal L}_0\tilde{\rho}+\frac{1}{{\rm i}\hbar}[H_{\rm mec},\tilde{\rho}],
\\ 
\tilde{\cal L}_1\tilde{\rho}
&=&\frac{1}{{\rm i}\hbar}[xV_1,\tilde{\rho}],\\ 
\tilde{\cal L}_2\tilde{\rho}
&=&\frac{1}{{\rm i}\hbar}[x^2V_2,\tilde{\rho}] 
+\tilde{\cal K}_2\tilde{\rho}. \label{L:2}
\end{eqnarray}
Here, $V_1$, $V_2$ are the first and second order terms in the expansion of
$\tilde{V}$ and read
\begin{eqnarray}
\label{V1}
V_1&=&\frac{{\rm i}\hbar}{2}\sum_{j=1,2}k_j\cos\phi_j
\Omega_j\left(|e\rangle\langle g_j|
  -|g_j\rangle\langle e|\right),\\
V_2&=&-\frac{\hbar}{4}\sum_{j=1,2}k_j^2\cos^2\phi_j
\Omega_j\left(|e\rangle\langle g_j|  + |g_j\rangle\langle e|\right).
\end{eqnarray}
The Liouvillian $\tilde{\cal K}_2$ has the form:
\begin{equation}
\tilde{\cal K}_2\tilde{\rho}=\alpha \sum_{j=1,2}\gamma_j k_j^2
|g_j\rangle\langle e|\left(2x\tilde{\rho} x
-x^2\tilde{\rho}-\tilde{\rho} x^2\right)|e\rangle\langle g_j|,
\end{equation}
where $\alpha=\int_{-1}^1{\rm d}\cos\phi{\cal N}(\cos\phi)\cos^2\phi$.

At zero order in $\zeta$, internal and external degrees of freedom are
decoupled: 
The state $\tilde{\rho}_{\rm St}$, solution of 
$\tilde{\cal L}_0\tilde{\rho}=0$,
is not uniquely defined, and has the form $\tilde{\rho}_{\rm St}=\rho_{\rm
  St}\otimes \mu(0)$, where $\rho_{\rm St}=\rho_{\rm D}$ is the internal steady
state and $\mu(0)={\rm Tr}_{\rm int}\{{\cal P}_0\tilde{\rho}(0)\}$ 
is the reduced density matrix, calculated from $\tilde{\rho}$ at $t=0$ by
tracing over the internal degrees of freedom (${\rm Tr}_{\rm int}\{\}$) and
applying the projector ${\cal P}_0$ acting over the external degrees of
freedom. The latter is defined as ${\cal
  P}_0\tilde{\rho}=\sum_{\epsilon}
\sum_{\psi_{\epsilon},\psi_{\epsilon}^{\prime}}^{\prime} 
|\psi_{\epsilon}\rangle\langle \psi_{\epsilon}^{\prime}| \langle
\psi_{\epsilon}|\tilde{\rho}|\psi_{\epsilon}^{\prime}\rangle$, where
$|\psi_{\epsilon}\rangle$, $|\psi_{\epsilon}^{\prime}\rangle$ are eigenstates
of $H_{\rm mec}$ at $\epsilon$. In general, at zero order equation
$\partial_t \tilde{\rho}=\tilde{\cal L}_0\tilde{\rho}$ admits an 
infinite number of stable solutions. They can be expanded in the basis of 
(left)
eigenvectors $\tilde{\rho}_{\epsilon,\epsilon^{\prime}}=\rho_{\rm St}\otimes
|\psi_{\epsilon}\rangle\langle\psi_{\epsilon^{\prime}}|$ 
at the (imaginary) eigenvalues
$\lambda_{\epsilon,\epsilon^{\prime}}=-{\rm
  i}(\epsilon-\epsilon^{\prime})/\hbar$ of the Liouville
operator $\tilde{\cal L}_0$, satisfying the secular equation 
$\tilde{\cal L}_0\tilde{\rho}_{\epsilon,\epsilon^{\prime}}
=\lambda_{\epsilon,\epsilon^{\prime}}
\tilde{\rho}_{\epsilon,\epsilon^{\prime}}$ 
($\tilde{\rho}_{\rm St}$ is eigenvector at $\lambda=0$). The eigenspaces 
at the eigenvalues  $\lambda_{\epsilon,\epsilon^{\prime}}$ may be also 
infinitely degenerate, as
it occurs for instance in the harmonic oscillator. For
$\zeta\neq 0$ these subspaces are coupled by $\tilde{\cal L}_1$, 
$\tilde{\cal L}_2$.
At second-order perturbation theory in $\zeta$, for
$\zeta_j\Omega_j\ll \min_{\epsilon,\epsilon^{\prime}\neq \epsilon}
\left(|\epsilon-\epsilon^{\prime}|\right)$ (i.e. when the spectum of
$\tilde{\cal L}_0$ is 
sufficiently spaced, to allow for non-degenerate perturbation theory), 
a closed equation for the dynamics in the subspace at $\lambda=0$ can be
derived. Denoting with $\tilde{\cal P}_0$ the projector onto this subspace,
defined as $\tilde{\cal P}_0\tilde{\rho}=\rho_{\rm St}\otimes {\rm Tr}_{\rm
  int}\{{\cal P}_0\tilde{\rho}\}$,  
this equation has the form~\cite{Lindberg84}:
\begin{equation} \label{Master:Red}
\frac{\rm d}{{\rm d}t}\tilde{\cal P}_{0}\tilde{\rho}(t)
=\left[\tilde{\cal P}_{0}\tilde{\cal L}_2\tilde{\cal P}_{0}+
  \int_{0}^{\infty}{\rm d}\tau\tilde{\cal P}_{0}
\tilde{\cal L}_1 {\rm e}^{\tilde{\cal
      L}_0\tau}\tilde{\cal L}_1\tilde{\cal P}_{0}\right]
\tilde{\rho}(t).
\end{equation}
After substituting the explicit form of 
$\tilde{\cal L}_1,\tilde{\cal L}_2$ in the second term on the
right-hand side of (\ref{Master:Red}) and tracing over the internal degrees of
freedom, we obtain:
\begin{eqnarray}
\frac{\rm d}{{\rm d}t}\mu &=&-{\cal P}_0\frac{1}{\hbar^2}\int_0^{\infty}{\rm
  d}\tau\Bigl( \mbox{Tr}_{\rm int} \left\{V_1{\rm e}^{{\cal
      L}_0\tau}V_1\rho_{\rm St}\right\} [\hat{x},[\hat{x}(\tau),\mu]] 
\nonumber\\
&+&\mbox{Tr}_{\rm int} \left\{V_1{\rm e}^{{\cal
      L}_0\tau}[V_1,\rho_{\rm St}]\right\}
[\hat{x},\mu\hat{x}(\tau)]\Bigr).\label{Master:1a}
\end{eqnarray} 
Here, the matrix $\mu={\rm Tr}_{\rm int}\{{\cal P}_0\tilde{\rho}\}$ 
is the reduced density matrix for the external degrees of freedom in the
subspace at eigenvalue (at zero order) $\lambda=0$.
The operator $\hat{x}(\tau)$ is here defined as
$\hat{x}(\tau)=\exp(-{\rm i}H_{\rm mec}\tau/\hbar)\hat{x}
\exp({\rm i}H_{\rm mec}\tau/\hbar)$. 

It is remarkable that the term 
$\tilde{\cal P}_{0}\tilde{\cal L}_2\tilde{\cal P}_{0}=0$. 
This result is explained by looking at the form of (\ref{L:2}).
When tracing over the internal degrees of freedom, the first
term of (\ref{L:2}) gives rise 
to a contribution proportional to ${\rm Tr}_{\rm int}\{V_2\rho_{\rm St}\}$:
This term usually gives rise to a shift to the
eigenvalues $\lambda_{\epsilon,\epsilon^{\prime}}$, it represents 
a renormalization of the harmonic oscillator frequency due to the
presence of the laser fields, and here
it vanishes since there is no occupation of the
excited state at steady state. The second term in
(\ref{L:2}) describes the diffusion arising from spontaneous emission 
into other mechanical states~\cite{Cirac92}. 
Again, since at steady state there is no excited-state
occupation, it vanishes. Thus, the disapperance of 
$\tilde{\cal P}_0\tilde{\cal L}_2\tilde{\cal P}_0$
is due to quantum interference at zero order in the Lamb-Dicke expansion.

For a non-degenerate spectrum of eigenvalues $\epsilon$, the
reduced matrix $\mu$ is diagonal, and the equation for a matrix element has
the form  
\begin{equation}
\label{Master:2}\frac{\rm d}{{\rm d}t}
\langle\psi_{\epsilon}|\mu|\psi_{\epsilon}\rangle
=\sum _{\epsilon^{\prime}}
C_{\epsilon,\epsilon^{\prime}} S(\omega_{\epsilon,\epsilon^{\prime}})
\left[-\langle \psi_{\epsilon}|\mu |\psi_{\epsilon}\rangle+
\langle \psi_{\epsilon^{\prime}}|\mu|\psi_{\epsilon^{\prime}}
\rangle\right]+{\rm
  H.c},
\end{equation}
where the coefficient $S(\omega_{\epsilon,\epsilon^{\prime}})$ is the value of
fluctuation spectrum of the operator $V_1$ at the frequency
${\omega_{\epsilon,\epsilon^{\prime}} =
\left[\epsilon-\epsilon^{\prime}\right]/\hbar}$, and reads
\begin{equation}\label{Snu}
S(\omega_{\epsilon,\epsilon^{\prime}})=\frac{1}{\hbar^2}\int_0^{\infty}{\rm
  d}\tau {\rm Tr}_{\rm int} \left\{V_1{\rm e}^{{\cal
      L}_0\tau}V_1\rho_{\rm St}\right\}{\rm e}^{{\rm
    i}(\epsilon-\epsilon^{\prime})\tau/\hbar}.
\label{Som}
\end{equation}
The coefficient $C_{\epsilon,\epsilon^{\prime}} =
|\langle\psi_{\epsilon}|x|\psi_{\epsilon^{\prime}}\rangle|^2$
weights the coupling between the center-of-mass states 
$|\psi_{\epsilon}\rangle$ and
$|\psi_{\epsilon^{\prime}}\rangle$ due to the photon momentum at
second order in the Lamb-Dicke expansion.
The equations necessary for the derivation of the explicit form 
of (\ref{Som}) are reported in appendix A. Equation~(\ref{Som}) shows
that the rate for
the transition $|\psi_{\epsilon}\rangle\to|\psi_{\epsilon^{\prime}}\rangle$
is given by 
the value of the excitation spectrum for a probe, whose interaction with the
atomic transition is described by $V_1$ and which is detuned 
from the pump by  $\omega_{\epsilon,\epsilon^{\prime}}$ (sideband transition). 
Here, the form of the potential enters explicitly through the
coefficients $C_{\epsilon,\epsilon^{\prime}}$, and implicitly through the
assumptions on the spectrum that have lead to~(\ref{Master:2}). 

\subsection{Harmonic oscillator}

We now let the potential be harmonic at frequency $\nu$,
$U(x)=\frac{1}{2}m\nu^2x^2$, and introduce the annihilation and creation
operators $a$ and $a^{\dagger}$ of a quantum of vibrational energy $\hbar\nu$,
such that $x=x_0(a^{\dagger}+a)$, $p={\rm i} p_0(a^{\dagger}-a)$, with
$x_0=\sqrt{\hbar/2m\nu}$ and $p_0=\sqrt{\hbar m\nu/2}$. The center-of-mass
Hamiltonian reads
\begin{equation}
H_{\rm mec}=\hbar\nu \left(a^{\dagger}a+\frac{1}{2}\right).
\end{equation}
Now, $|\psi_{\epsilon}\rangle=|n\rangle$ and $\epsilon=\hbar\nu(n+1/2)$, where
$n=0,1,\ldots$ is the number of phonon excitations, and the mechanical energies
are equidistantly spaced by $\hbar\nu$. The coefficients
$C_{n,n^{\prime}}=x_0^2(n\delta_{n^{\prime},n-1}+(n+1)
\delta_{n^{\prime},n+1})$, and thus at first order in the Lamb-Dicke expansion
the relevant transitions between motional states are
the blue sideband $|n\rangle\to|n+1\rangle$ at frequency
$\omega_L-\nu$,  and the red sideband
$|n\rangle\to|n-1\rangle$ at frequency $\omega_L+\nu$. We define the
Lamb-Dicke parameter $\eta_j=k_jx_0$, that fulfills the relation
$\zeta_{1,2}=\eta_{1,2}\sqrt{2\langle n\rangle+1}$, with $\langle n\rangle$
average number of phonon excitations. \\ 
For the harmonic oscillator the equations derived in the
previous section simplify notably: Equation
(\ref{Master:1a}) gets the form
\begin{eqnarray}
\label{Master:3}
\frac{\rm d}{{\rm d}t}\mu &=& x_0^2 S(\nu)\left[-a^{\dagger}a\mu+a\mu
  a^{\dagger}\right]\nonumber\\ &+& x_0^2
S(-\nu)\left[-aa^{\dagger}\mu+a^{\dagger}\mu a\right]+{\rm H.c},
\end{eqnarray}
and for the probability $\mu_{n,n}=\langle n|\mu|n\rangle$ of the system to be
in the number state $|n\rangle$, 
Eq.~(\ref{Master:3}) turns to a rate equation, whose form is well known
in laser cooling of single ions~\cite{Stenholm86},
\begin{eqnarray} \label{Rate}
\frac{\rm d}{{\rm d}t}\mu_{n,n}
  &=&\eta^2[(n+1)(A_-\mu_{n+1,n+1}-A_+\mu_{n,n})\nonumber\\
  & &+ n (A_-\mu_{n,n}-A_+\mu_{n-1,n-1})].
\end{eqnarray}
In our case of a three-level atom, 
$\eta=\eta_1\cos\phi_1-\eta_2\cos\phi_2$, and 
\begin{equation}
\label{Snu:2} A_{\pm}=2\mbox{Re}[S(\mp \nu)]=\frac{1}{4}
\left(\frac{\Omega_1\Omega_2}{\Omega}\right)^2 \frac{\gamma\nu^2}
{\left[\Omega^2/4-\nu(\nu\pm\Delta)\right]^2+ \gamma^2\nu^2/4}.
\end{equation}
Equation~(\ref{Rate}) has the same structure as the rate equation derived for 
sideband cooling in a two-level system. 
Here, however, the rates $A_{\pm}$ describe the sideband
excitation including the effect of quantum interference between the atomic
transitions. Equation~(\ref{Rate})
allows for a steady state when $A_->A_+$, which is fulfilled when $\Delta<0$
(blue detuning) and $\Omega>2\nu$, or when $\Delta>0$ (red detuning) and
$\Omega<2\nu$. The value of the trap frequency $\bar{\nu}=\Omega/2$ separates
two regimes: for $\nu<\Omega/2$ it is the narrow resonance that determines
relevantly the center-of-mass dynamics, whereas for $\nu>\Omega/2$ the
sideband transitions are at the frequency range of the broad
resonance~\cite{Footnote}. We remark that 
$\eta=0$ for $k_1\cos\phi_1=k_2\cos\phi_2$, corresponding to the Doppler-free
situation. Furthermore, the Lamb-Dicke parameter entering into the dynamics is
the one determined by the laser wave vector. 
The Lamb-Dicke parameter connected to
spontaneous emission events (i.e. recoils because of emission into other states
of the motion) does not appear, since the diffusion term vanishes at second
order in the Lamb-Dicke expansion.\\
\indent In the following we assume $k_1\cos\phi_1\neq k_2\cos\phi_2$ and
$\Delta<0$, $\Omega>2\nu$ ($A_->A_+$). Insight into the dynamics
can be gained from the equation for the
average number of phonon $\langle n(t)\rangle = \sum_{n=0}^{\infty}n\mu_{n,n}
(t)$,
that is derived from~(\ref{Rate}) and has the form~\cite{Stenholm86}
\begin{equation}
\label{n:t}
\frac{\rm d}{{\rm d}t}\langle n\rangle=-W\langle n\rangle + 
\eta^2 A_+,
\end{equation}
where $W=\eta^2(A_--A_+)$ is the cooling rate.  
The steady state value $\langle n\rangle_{\infty}$ reads
\begin{equation}
\langle n\rangle_{\infty}
=\frac{4\left[\Omega^2/4-\nu(\nu-\Delta)\right]^2 +
 \gamma^2\nu^2}{4\nu|\Delta|(\Omega^2-4\nu^2)},
\end{equation}
and is minimum when $\Omega^2=4\nu(\nu-\Delta)$. This relation 
corresponds to setting the a.c. Stark shift $\delta\omega_+$ of the 
narrow resonance $|\Psi_+\rangle$ at the frequency of the first red sideband,
$\delta\omega_+=\Delta-\nu$. For this value, 
$\langle n\rangle_{\infty}^{({\rm min})}=(\gamma/4|\Delta|)^2$:
Hence, low temperatures are achieved for lasers far detuned from 
atomic resonance. This corresponds to an enhanced asymmetry of the excitation
spectrum, as the one shown in Fig.~2, where the two resonances have very
different widths.

For $\delta\omega_+=\Delta-\nu$ the cooling rate scales as 
\begin{equation}
W^{\rm max}\sim \eta^2 (\Omega_1\Omega_2/\Omega)^2/\gamma. 
\end{equation}
Thus, fast cooling is
achieved for large Rabi frequencies and when $\Omega_1=\Omega_2$.
The ultimate limit to $W$ is set by the parameters that ensure the 
validity of the perturbative treatment here applied: This is valid
for $\eta_j\Omega_j\cos\theta \ll \gamma_+$ ($j=1,2$), 
with $\gamma_+\sim\gamma\cos^2\theta$ linewidth of
the narrow resonance, corresponding in the bare atom to the situation where
the probe (the sideband) does not saturate the transition to 
$|\Psi_+\rangle$. At $\delta\omega_+=\Delta-\nu$, one has 
$\gamma_+\sim\nu\gamma/4\sqrt{\Delta^2+\Omega^2}$, which sets
the fastest rate at which efficient laser cooling can occur, 
$W^{\rm max}\sim \gamma_{+}/2$.

It is remarkable that these results do not depend on the branching ratio
$\gamma_1/\gamma_2$. In fact, in this limit the branching ratio
enters the problem only through $T_0$. 
Nevertheless, a too large branching ratio affects the time scale $T_0$
at which the transient steady-state is reached. 

In Fig. 3 we test the validity of the adiabatic elimination procedure for
various values of the Lamb-Dicke parameter, by comparing the results predicted
by (\ref{n:t}) with a full numerical simulation. The parameters are
reported in the caption. Full  agreement between the two results is
found for $\eta=0.02$ ($\eta_1\cos\phi_1=-\eta_2\cos\phi_2=0.01$). 
It should be mentioned that
in~\cite{EITtheo2000} full agreement has been found for
$\eta$ as large as 0.2. 
On the other hand, those results have been evaluated for the case
$\Omega_1\ll \Omega_2$, and the small value of $\Omega_1$ 
ensured the validity of the perturbative expansion.
\begin{center}
\begin{figure}
\epsfxsize=0.6\textheight
\epsffile{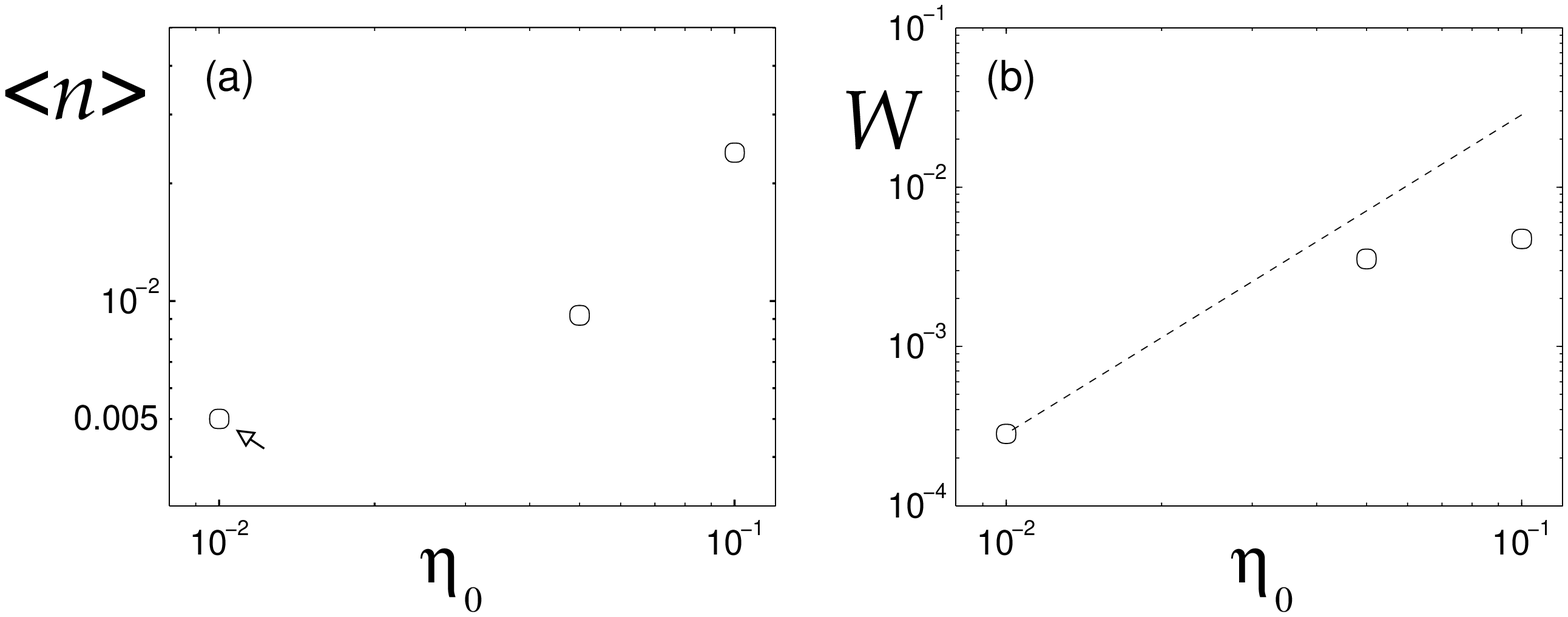}
\caption{Plot of (a) $\langle n\rangle_{\infty}$ and (b)
  $W$ as a function of $\eta_0$. Dashed line: rate equation 
 result; 'o':
 numerical simulation, 500 trajectories with the Quantum Monte Carlo method. 
Here, $\eta_1\cos\phi_1=\eta_2\cos\phi_2=\eta_0$ ($\eta=2\eta_0$), 
$\nu=2$~MHz, 
$\gamma=20$~MHz, $\Omega_1=\Omega_2=17$~MHz,
$\Delta=70$~MHz, $\gamma_1/\gamma_2=1$. In (a) the result of the simulation in
 agreement with the rate equation prediction
 ($\langle n\rangle=0.005$) is indicated by the arrow. 
In (b) the rate $W$ is in units of $\gamma/2$. }
\end{figure}
\end{center}

\subsubsection{Discussion}

We have shown that, by properly choosing the lasers parameters, 
one can achieve almost unity ground-state occupation with this cooling method
(EIT cooling). The state $|\psi_{\rm D}\rangle|0\rangle$ 
is equivalent to the ground state in sideband cooling, since it is only 
off-resonantly (weakly) coupled to other states, and it satisfies the 
criteria of an approximate dark state as discussed in~\cite{Dum94}.

From~(\ref{Snu:2}) one recovers the rates of Eq.~(4) 
in~\cite{EITtheo2000} in the limit $\Omega_1\ll\Omega_2$. 
We have shown that the same dynamics are
encountered in more general situations, that do not impose a specific relation
between the two Rabi frequencies.
From the technical point of view EIT cooling proves again to be more
advantageous than Raman sideband cooling (see~\cite{EITtheo2000,ApplB2001}). 
Such advantage is mainly twofold. On one hand, in EIT cooling both lasers
cool  
the atom, and a decay into one or the other channel does not affect the
efficiency of the process, while in Raman sideband cooling a finite branching
ratio gives rise to heating~\cite{OptComm2000}. Another important feature of
EIT cooling is
the disappearance of the carrier absorption due to quantum interference.
This effect implies the suppression of
diffusive processes: Since in the coarse-grained evolution the excited state is
effectively empty, processes, where the atom is scattered into other motional
states by spontaneous emission, disappear at second order in the Lamb-Dicke
expansion. That implies an improved efficiency with respect to Raman-sideband
cooling, where instead such processes are present, as already discussed 
in~\cite{EITtheo2000}. 

It is instructive to compare the dynamics in EIT cooling with the dynamics 
of a trapped ion at the node of a standing wave, as studied for example 
in~\cite{Cirac92}.
At the node of a standing wave the carrier absorption cancels, since here the
value of the electric field is zero. Nevertheless, 
sideband absorption occurs because of the finite size of
the motional wave packet. In the case of a
$\Lambda$- configuration driven by two travelling waves 
at two-photon resonance, the transient
dark state~(\ref{Dark:1}) is a superposition of the
states $|g_1\rangle$ and $|g_2\rangle$ whose relative phase
is a function of the coordinate $x$, so that the finite size of the wave
packet allows sideband absorption also in this case. 
Nevertheless, in the LDR the gradient of the phase over the wavepacket
is small, and the sideband transitions 
are excited on a longer time scale. This can be illustrated when writing the
atom-laser interaction~(\ref{Vmec}) at the first order in the Lamb-Dicke
expansion and in the form
\begin{equation}
\label{Vapprox}
\tilde{V}\approx
\hbar\frac{\Omega}{2}
\left[|e\rangle\langle \Psi_{\rm C}|+{\rm i} |e\rangle\langle
  \Psi_{\rm D}|kx +{\rm H.c.}\right],
\end{equation}
where we have made the simplifying assumptions
$\Omega_1=\Omega_2=\Omega/\sqrt{2}$, $k_1\cos\phi_1=-k_2\cos\phi_2=k$. 
Here, we see that the dark state is coupled to the excited state at first
order in the Lamb-Dicke expansion, for effects arising from the
finite size of the motional wave packet. 

The atom dynamics during the coarse-grained evolution can be interpreted in
terms of field gradients over the size of the wave-packet, that give rise to
forces~\cite{Nienhuis91}. In this respect, one can say that this method uses
the phase gradient of the dark state, due to the spatial gradient of the
total field, for achieving cooling. In this
context, we remark that the operator $V_1$ in~(\ref{Snu}) is the gradient
of the potential~(\ref{Vmec}) at $x=0$, i.e. at the center of trap.

Finally, we apply the results obtained for the harmonic oscillator
to the case of a generic potential $U(x)$. Several conclusions drawn 
in this section are applicable to the case described in Eq.~(\ref{Master:2}), 
when the mechanical Hamiltonian has a discrete spectrum, and the minimum
distance between two neighbouring energy level is sufficiently large to allow
for non-degenerate perturbation theory. 
Laser cooling is here achieved for the same parameters as for the harmonic
oscillator. 
However, the narrow resonance enhances transitions in a finite range
of frequencies ($<\bar{\nu}$), 
and $\delta\omega_+$ must be properly tuned, e.g. to the
average value of the red sideband transitions frequencies. 
The process will thus be efficient under the condition that, for each motional
state, there is a sufficient number of red sidebands inside this range, so
that the rate of cooling for a given motional state 
is larger than the rate of heating.

An interesting question is how the dynamics are affected when
the external potential depends on the electronic state, and thus when
\begin{eqnarray*}
U(x) = U_1(x)|g_1\rangle\langle g_1| + U_e(x)|e\rangle\langle e| +
U_2(x)|g_2\rangle\langle g_2|.
\end{eqnarray*}
We consider first the case $U_1(x)=U_2(x)$, while $U_e(x)$ is -say- constant,
so that the center of mass of the excited atom is not spatially confined and
the spectrum of $H_{\rm mec}$ at the state $|e\rangle$ is a continuum.
Assuming that for $U_1$, $U_2$ the Lamb-Dicke regime holds, then at two-photon
resonance and
during the transient dynamics the atom is optically pumped into the
(transient) dark state~(\ref{Dark}). However, during $T_0$ the center-of-mass
wave packet changes, since each eigestate of $U_1(x)$ ($U_2(x)$) may
have non-zero
overlap with several eigenstates of $U_e(x)$. This effect constitutes a
diffusion mechanism, that lowers the cooling efficiency and,
outside of some regimes, can make it even impossible. Formally, for
$U_e\neq U_1,U_2$, the formalism applied in this section is not
applicable, since one cannot separate the time scales characterizing the
evolution of the internal and external degrees of freedom. \\
In the general case of three different confining potentials
the presence of a dark state cannot be excluded: that however depends on
the specific form of the functions $U_j(x)$.

\section{Conclusions and outlook}

We have presented a systematic investigation of the center-of-mass
dynamics of a trapped ion, the internal transitions of which are driven by
lasers in a $\Lambda$-type configuration and set at two-photon resonance. 
Assuming that the center-of-mass wavepacket is well localized over the laser
wavelength (Lamb-Dicke regime), we have adiabatically eliminated the
internal degrees of freedom from the equation
of the center-of-mass dynamics, and obtained a set of rate equations for the
occupation of the motional states. We have identified the parameter regimes
where efficient ground-state cooling can be achieved. The derivation 
here presented provides the theoretical background 
for the equations in~\cite{EITtheo2000,EITexp2000,ApplB2001} and extends the
parameter regime to cases which have not been previously considered. 
As also discussed in~\cite{EITtheo2000}, we have shown that 
diffusive processes, encountered in cooling with two-level atoms or
with effective two-level systems (Raman sideband cooling), 
are suppressed because of quantum interference between the
dipole transitions at zero order in the Lamb-Dicke expansion.
Cooling takes place because of excitations due to the spatial
gradient of the electric field over the width of the motional
wave-packet, that are due to the finite size of the wave-packet itself and
occur at first order in the Lamb-Dicke expansion. The motion can be said to be
cooled by both lasers, while the branching ratio does not affect in general
the efficiency of the process. \\
Finally, 
we have discussed the possibility to observe these dynamics for other types
of potentials, that may depend on the electronic state.

This work opens interesting prospects in the manipulation 
of the quantum center-of-mass motion of atoms by using quantum
interference in driven multilevel transitions, that is subject of on-going
investigations.

\section{Acknowledgements}

The author ackowledges J\"urgen Eschner, who has encouraged and stimulated the
completion of this work with several discussions and critical comments.
He is also ackowledged for the critical reading of
this manuscript. Most part of this work has been done at the
Max-Planck-Institut f\"ur Quantenoptik: the author is grateful to
 Herbert Walther for his support, and to the whole group in Garching for 
the enjoyable and stimulating scientific atmosphere. The author thanks 
Markus Cirone for the corrections to the english.

\begin{appendix}

\section{Calculation of $S(\nu)$}

The term $S(\nu)$ in~(\ref{Snu}) is the Laplace transform at ${\rm i}\nu$
of the correlation function $G(\tau)$, defined as
$G(\tau)={\rm Tr}_{\rm int}\{V_1(\tau) V_1(0)\rho_{\rm St}\}$, 
where $V_1(\tau)=V_1{\rm e}^{{\cal L}_0\tau}$. This is evaluated applying 
the quantum regression theorem~\cite{AtomPhoton,Narducci90}. 
In the following, we derive
the equations that are essential for this calculation. For convenience, we
introduce the vector-operator $\hat{\sigma}$ whose components are defined as:
$\hat{\sigma}_1=|g_1\rangle\langle g_1|$, $\hat{\sigma}_2=|g_2\rangle\langle
g_2|$, $\hat{\sigma}_3=|g_1\rangle\langle e|$,
$\hat{\sigma}_4=|e\rangle\langle g_1|$, $\hat{\sigma}_5=|g_2\rangle\langle
e|$, $\hat{\sigma}_6=|e\rangle\langle g_2|$,
$\hat{\sigma}_7=|g_2\rangle\langle g_1|$, $\hat{\sigma}_8=|g_1\rangle\langle
g_2|$. The mean value $\langle \hat{\sigma}_j\rangle={\rm
  Tr}\{\hat{\sigma}_j\rho\}$ obeys the equations ${\rm d}\langle
\hat{\sigma}_j\rangle/{\rm d}t=M\langle \hat{\sigma}_j\rangle+B$, where
$M$, $B$ are a matrix and a column vector, respectively, and are defined
through the equations
\begin{eqnarray}
&&\sum_{j=1}^8 M_{1,j}\langle\hat{\sigma}_j\rangle
=-\gamma_1(\langle\hat{\sigma}_{1}\rangle+
\langle\hat{\sigma}_{2}\rangle)
-{\rm i}\frac{\Omega_1}{2}\left(
\langle\hat{\sigma}_3\rangle-
\langle\hat{\sigma}_4\rangle\right),
\nonumber\\
&&\sum_{j=1}^8 M_{2,j}\langle\hat{\sigma}_j\rangle
=-\gamma_2(\langle\hat{\sigma}_{1}\rangle+
\langle\hat{\sigma}_{2}\rangle)
-{\rm i}\frac{\Omega_2}{2}\left(
\langle\hat{\sigma}_{5}\rangle-
\langle\hat{\sigma}_{6}\rangle\right),
\nonumber\\
&&\sum_{j=1}^8 M_{3,j}\langle\hat{\sigma}_j\rangle
=-{\rm i}\frac{\Omega_1}{2}\left(2\langle\hat{\sigma}_1\rangle+
\langle\hat{\sigma}_2\rangle\right)
-\left(\frac{\gamma}{2}+{\rm i}\Delta\right)\langle\hat{\sigma}_3\rangle
-{\rm i}\frac{\Omega_2}{2}\langle\hat{\sigma}_8\rangle,
\nonumber\\
&&\sum_{j=1}^8 M_{4,j}\langle\hat{\sigma}_j\rangle
={\rm i}\frac{\Omega_1}{2}\left(2\langle\hat{\sigma}_1\rangle+
\langle\hat{\sigma}_2\rangle\right)
-\left(\frac{\gamma}{2}-{\rm i}\Delta\right)\langle\hat{\sigma}_4\rangle
+{\rm i}\frac{\Omega_2}{2}\langle\hat{\sigma}_7\rangle,
\nonumber\\
&&\sum_{j=1}^8 M_{5,j}\langle\hat{\sigma}_j\rangle=
-{\rm i}\frac{\Omega_2}{2}\left(\langle\hat{\sigma}_1\rangle+
2\langle\hat{\sigma}_2\rangle\right)
-\left(\frac{\gamma}{2}+{\rm i}\Delta\right)\langle\hat{\sigma}_5\rangle
-{\rm i}\frac{\Omega_1}{2}\langle\hat{\sigma}_7\rangle,
\nonumber\\
&&\sum_{j=1}^8 M_{6,j}\langle\hat{\sigma}_j\rangle=
{\rm i}\frac{\Omega_2}{2}\left(\langle\hat{\sigma}_1\rangle+
2\langle\hat{\sigma}_2\rangle\right)
-\left(\frac{\gamma}{2}-{\rm i}\Delta\right)\langle\hat{\sigma}_6\rangle
+{\rm i}\frac{\Omega_1}{2}\langle\hat{\sigma}_8\rangle,
\nonumber\\
&&\sum_{j=1}^8 M_{7,j}\langle\hat{\sigma}_j\rangle=
+{\rm i}\frac{\Omega_2}{2}\langle\hat{\sigma}_4\rangle
-{\rm i}\frac{\Omega_1}{2}\langle\hat{\sigma}_5\rangle,
\nonumber\\
&&\sum_{j=1}^8 M_{8,j}\langle\hat{\sigma}_j\rangle=
-{\rm i}\frac{\Omega_2}{2}\langle\hat{\sigma}_3\rangle
+{\rm i}\frac{\Omega_1}{2}\langle\hat{\sigma}_6\rangle,
\nonumber
\end{eqnarray}
and $B_j=\gamma_1\delta_{j,1}+\gamma_2\delta_{j,2}+
{\rm i}\frac{\Omega_1}{2}(\delta_{j,3}-\delta_{j,4})+
{\rm i}\frac{\Omega_2}{2}(\delta_{j,5}-\delta_{j,6})$,
with $j=1,\ldots,8$  and $\delta_{j,k}$ the Kronecker-delta.
According to this definition, the steady-state vector is now
$\sigma_{\rm St}=M^{-1}B$.\\
Using this notation, we rewrite the operator $V_1$ in~(\ref{V1})
as $V_1=\alpha_1(\hat{\sigma}_4-\hat{\sigma}_3)+
\alpha_2(\hat{\sigma}_6-\hat{\sigma}_5)$, with
$\alpha_j={\rm i}\hbar k_j\cos\phi_j\Omega_j/2$, $j=1,2$. The 
Laplace transform $S(\nu)$ is then the sum of the Laplace
transforms $s_j(\nu)$ of the individual terms 
$g_j(\tau)={\rm Tr}\{\hat{\sigma}_j(\tau)V_1(0)\sigma_{\rm St}\}$,
such that $S(\nu)=\alpha_1(s_4(\nu)-s_3(\nu))+\alpha_2
(s_6(\nu)-s_5(\nu))$, where $s_j(\nu)$  are given by the equations
\begin{eqnarray*}
s_j(\nu)=\sum_kL_{jk}
\left({\rm Tr}\{\sigma_kV_1(0)\rho_{\rm St}\}
+\frac{1}{{\rm i}\nu}B_k{\rm Tr}\{V_1(0)\rho_{\rm St}\}\right),
\end{eqnarray*}
with $L$ matrix, $L=[{\rm i}\nu-M]^{-1}$.

\end{appendix}


\begin{references}

\bibitem{Metcalf}
H.J.~Metcalf and P.~van~der~Straten, {\it Laser Cooling and Trapping},
Springer-Verlag, New York (1999).

\bibitem{SidebandIons}
F.~Diedrich, J.C.~Bergquist, W.M.~Itano, and D.J.~Wineland,
Phys. Rev. Lett. {\bf 62}, 403 (1989); Ch.~Roos, Th.~Zeiger, H.~Rohde,
H.C.~N\"agerl, J.~Eschner, D.~Leibfried, F.~Schmidt-Kaler,
and R.~Blatt, Phys. Rev. Lett. {\bf 83}, 4713 (1999).

\bibitem{RamanIons}
C.~Monroe, D.M.~Meekhof, B.E.~King, S.R.~Jefferts, W.M.~Itano, D.J.~Wineland,
and P.~Gould, Phys. Rev. Lett. {\bf 75}, 4011 (1995); 

\bibitem{RamanAtoms}
M.~Morinaga, I.~Bouchoule, J.-C.~Karam, and C.~Salomon, Phys. Rev. Lett. {\bf
  83}, 4037 (1999); A.J.~Kerman, V.~Vuletic,
C.~Chin, and S.~Chu, Phys. Rev. Lett. {\bf 84}, 439 (2000);
D.-J.~Han, S.~Wolf, S.~Oliver, C.~McCormick, M.T.~DePue, and D.S.~Weiss,
Phys. Rev. Lett. {\bf 85}, 724 (2000).

\bibitem{QC}
The proposals are several, and it goes beyond the scope of this paper to give
a complete bibliography. Some, currently inspiring experiments with
trapped ions and neutral atoms in optical lattices, are, respectively:  
J.I.~Cirac and P.~Zoller, Nature (London) {\bf 404}, 579 (2000);  
D. Jaksch, H.-J. Briegel, J.I.~Cirac, C.W.~Gardiner, and P.~Zoller,
Phys. Rev. Lett. {\bf 82}, 1975 (1999). 

\bibitem{EITtheo2000}
G. Morigi, J. Eschner, C.H. Keitel, Phys. Rev. Lett. {\bf 85}, 4458 (2000).

\bibitem{CPT}
E.~Arimondo, {\it Progress in Optics XXXV}, ed. by E.~Wolf (North-Holland,
Amsterdam, 1996), pp. 259-354; S.E.~Harris, Phys. Today {\bf 50}, No. 7, 36 
(1997).

\bibitem{EITexp2000}
C.F.~Roos, D.~Leibfried, A.~Mundt, F.~Schmidt-Kaler, J.~Eschner, and R.~Blatt,
Phys. Rev. Lett. {\bf 85}, 5547 (2000).


\bibitem{ApplB2001}
F.~Schmidt-Kaler, J.~Eschner, G.~Morigi, C.~Roos, D.~Leibfried, A.~Mundt, and
R.~Blatt, Appl. Phys. B {\bf 73}, 807 (2001).

\bibitem{Hansch2001}
F.~Schmidt-Kaler, J.~Eschner, R.~Blatt, D.~Leibfried, C.~Roos,
G.~Morigi, {\it Laser Cooling of Trapped Ions} in {\it Laser Physics at the
  Limits}, ed. by H. Figger, D. Meschede, C. Zimmerman, Springer-Verlag,
Berlin (2001). 

\bibitem{Lindberg84}
J. Javanainen, M. Lindberg, and S. Stenholm, J. Opt. Soc. Am. B {\bf 1}, 111
(1984); 
M. Lindberg and S. Stenholm, J. Phys. B: At. Mol. Phys. {\bf 17}, 3375
(1984).

\bibitem{Lindberg86}
M. Lindberg and J. Javanainen, J. Opt. Soc. Am. B {\bf 3}, 1008 (1986).

\bibitem{Wineland92}
D.J. Wineland, J. Dalibard, and C. Cohen-Tannoudji, J. Opt. Soc. Am. B {\bf
  9}, 32 (1989). 

\bibitem{Marzoli94}
I. Marzoli, J.I. Cirac, R. Blatt, and P. Zoller, Phys. Rev. A {\bf 49}, 2771
(1994).

\bibitem{Dum94}
R. Dum, P. Marte, T. Pellizzari, and P. Zoller, Phys. Rev. Lett. {\bf 73},
2829 (1994).

\bibitem{AtomPhoton}
C. Cohen-Tannoudij, J. Dupont-Roc, G. Grynberg, {\it Atom-Photon
Interactions}, J. Wiley and Sons ed. (Toronto, 1992).

\bibitem{Lounis92}
B. Lounis and C. Cohen-Tannoudij, J. de Phys. II (France) {\bf 2}, 579 (1992).

\bibitem{Janik85}
G.~Janik, W.~Nagourney, and H.~Dehmelt, J. Opt. Soc. of Am. {\bf 2}, 1251
(1985). 

\bibitem{Stalgies98}
Y. Stalgies, I. Siemers, B. Appasamy, and P.E. Toschek, J. Opt. Soc. Am. B
{\bf 15}, 2505 (1989). 

\bibitem{VSCPT}
The first experiment where velocity selective coherent population trapping  
has been observed is reported in 
A. Aspect, E. Arimondo, R. Kaiser, N. Vansteenkiste, and C. Cohen-Tannoudji,
Phys. Rev. Lett. {\bf 61}, 826 (1988) and J. Opt. Soc. Am. B {\bf 6}, 2112
(1989). For a review, see C.~Cohen-Tannoudji, {\it Nobel-Prize lectures},
Rev. Mod. Phys. {\bf 70}, 707 (1998) and \cite{Metcalf}.

\bibitem{Cirac92}
J.I. Cirac, R. Blatt, P. Zoller, and W.D. Phillips, Phys. Rev. A {\bf 46},
2668 (1992).

\bibitem{Footnote}
The frequency $\bar{\nu}=\Omega/2$ sets the upper bound to the 
modes that can be simultaneously cooled by means of this scheme, 
for example when it is applied
to cooling of all three motion axes of a trapped ions~\cite{EITexp2000} 
or to the axial modes of a linear ion chain~\cite{ApplB2001}.

\bibitem{Stenholm86}
S. Stenholm, Rev. Mod. Phys. {\bf 58}, 699 (1986).

\bibitem{OptComm2000}
G. Morigi, H. Baldauf, W. Lange, and H. Walther, Opt. Comm.  {\bf 187}, 171 
(2001).

\bibitem{Nienhuis91}
G.~Nienhuis, P.~van~der~Straten, S-Q.~Shang, Phys. Rev. A {\bf 44}, 462
(1991). 

\bibitem{Narducci90}
L.M. Narducci, M.O. Scully, G.-L. Oppo, P. Ru, and J.R. Tredicce, Phys. Rev. A
{\bf 42}, 1630 (1990); A.S. Manka, H.M. Doss, L.M. Narducci, P. Ru, and 
G.-L. Oppo, {\it ibid}. {\bf 43}, 3748 (1991).
\end{references}
\end{document}